\documentclass[a4paper,11pt]{article}

\pdfoutput=1

\usepackage{jcappub}

\title{A high precision semi-analytic mass function}

\author[a,b,c]{Antonino Del Popolo,}
\author[d]{Francesco Pace}
\author[e,f,g]{and Morgan Le Delliou}

\affiliation[a]{Dipartimento di Fisica e Astronomia, University Of Catania, \\
Viale Andrea Doria 6, 95125, Catania, Italy}
\affiliation[b]{INFN sezione di Catania,\\
Via S. Sofia 64, I-95123 Catania, Italy}
\affiliation[c]{International Institute of Physics, Universidade Federal do Rio Grande do Norte, \\
59012-970 Natal, Brazil}
\affiliation[d]{Jodrell Bank Centre for Astrophysics, School of Physics and Astronomy, The University of Manchester,\\
Manchester, M13 9PL, U.K.}
\affiliation[e]{Instituto de F\'isica Teorica, Universidade Estadual de S\~ao Paulo (IFT-UNESP),\\
 Rua Dr. Bento Teobaldo Ferraz 271, Bloco 2 - Barra Funda, 01140-070 S\~ao Paulo, SP Brazil}
\affiliation[f]{Institute of Theoretical Physics,
Physics Department, Lanzhou University,
No.222, South Tianshui Road, Lanzhou, Gansu 730000, P R China}
\affiliation[g]{Instituto de Astrof\'isica e Ci\^encias do Espa\c co, Universidade de Lisboa, Faculdade de Ci\^encias, 
Ed. C8, Campo Grande, 1769-016 Lisboa, Portugal}

\emailAdd{adelpopolo@oact.inaf.it}
\emailAdd{francesco.pace@manchester.ac.uk}
\emailAdd{delliou@ift.unesp.br}

\abstract{
In this paper, extending past works of Del Popolo, we show how a high precision mass function {
(MF)} can be obtained using the 
excursion set approach and an improved barrier taking implicitly into account a non-zero cosmological constant, the 
angular momentum acquired by tidal interaction of proto-structures and dynamical friction. 
In the case of the $\Lambda$CDM paradigm, we find {
that our MF is in agreement at the 3\% level to Klypin's Bolshoi simulation, in the  }  mass range {
$M_{\rm vir} = 5  \times 10^9 h^{-1} M_{\odot} \textnormal{ --- }  5 \times 10^{14} h^{-1} M_{\odot}$ and redshift range $0 \lesssim z \lesssim 10$.} For $z=0$ {
we also compared our MF to several fitting formulae, and found in particular agreement with Bhattacharya's within 3\% in the mass range $10^{12}-10^{16} h^{-1} M_{\odot}$. Moreover, we discuss our MF validity for different cosmologies.} 
}


\begin{document}

\maketitle

\flushbottom

\section{Introduction}
The $\Lambda$CDM model, often referred to as "cosmic concordance" model or standard model of Big Bang cosmology, is a 
"double dark" cosmological model, in which the Universe is constituted by cold dark matter plus a vacuum density 
energy, represented by the cosmological constant $\Lambda$. On large and intermediate scales this model has been proven 
to be very successful in fitting a large variety of data 
\citep{Spergel2003,DelPopolo2007,Komatsu2011,DelPopolo2013,Story2013,Das2014,DelPopolo2014,Planck2016_XIII}.
\footnote{
From a theoretical point of view, the model is afflicted by the fine tuning problem 
\citep{Weinberg1989,Astashenok2012} and the cosmic coincidence problem. At kpc-scales, the $\Lambda$CDM model is 
suffering other problems: the cusp/core problem \citep{Flores1994,Cardone2012,DelPopolo2012a,DelPopolo2012b,DelPopolo2013d,DelPopolo2014a} or the 
missing satellite problem \citep{Moore1999,DelPopolo2014}.}

A further fundamental test of the $\Lambda$CDM model resides in the accurate prediction of the halo mass function (MF), 
namely the mass distribution of dark matter halos, or more in detail the number density of dark matter halos per mass 
interval \citep[see][]{DelPopolo2007,Hiotelis2006,Hiotelis2013}.

At redshifts $z\leq 2$, the high mass end of the MF (clusters of galaxies) is very sensitive to cosmological 
parameters like $\sigma_8$, variations in cosmological parameters like the Universe matter and dark energy (DE) 
content ($\Omega_{\rm m}$ and $\Omega_{\Lambda}$), the equation of state parameter $w$ and its evolution 
\citep{Holder2001,Haiman2001,Weller2002,Majumdar2003,Pace2010,Pace2012,Pace2014,Malekjani2015,Naderi2015, 
NazariPooya2016}.
\footnote{$\sigma_8$ represents the linear power spectrum amplitude on a scale of 8 $h^{-1}$~Mpc.}

At higher redshifts, the MF is an important probe of the Universe reionization history 
\citep[e.g.][]{Furlanetto2006} and quasar abundance \citep[e.g.][]{Haiman2001a}.

Apart from its use to determine the cosmological parameters, the MF is a fundamental ingredient to study DM 
distribution, aspects of formation and evolution of galaxies through semi-analytic and analytic models. 
Furthermore, a high precision MF is related to ongoing and upcoming surveys detecting clusters using optical 
observations, X-rays, or the Sunyaev-Zel'dovich (SZ) effect. Thus, clearly, a simple and accurate high precision MF, 
valid for different cosmologies and redshifts and allowing a precise extraction of cosmological parameters is a helpful 
and valuable asset.

\cite{Press1974} (PS) proposed a simple model in which initial fluctuations are spherical, with a Gaussian 
distribution, and their evolution is followed from the linear phase until collapse using a spherical collapse model 
(SCM). At the virialization epoch (identified with the collapse redshift), the density contrast, 
$\delta=\frac{\rho-\bar{\rho}}{\bar{\rho}}$, calculated within linear perturbation theory, gets the value 
$\delta_{\rm c}\simeq1.686$ for an Einstein-de Sitter cosmology. Under the assumption that the density field has a 
Gaussian probability distribution, one can calculate the probability that the overdensity on a given scale exceeds the 
critical value $\delta_{\rm c}$, which is independent from the mass of the collapsing object. This quantity is 
proportional to the number of cosmic structures characterized by a density perturbation greater than $\delta_{\rm c}$. 
Unfortunately, in the PS theory the number of objects in the high mass tail is underpredicted, and conversely for 
objects in the low mass tail of the MF \citep[e.g.][]{Efstathiou1988,Gross1998,Jenkins2001,White2002}. Even the 
extended-PS formalism taking merging into account \citep{Bond1991,Bower1991,Lacey1993,Gardner2001} does not solve the 
problem.

As shown in \cite{DelPopolo1998} (Eq.~28, Fig.~6), the tidal interaction with neighbours and the angular momentum 
acquired modifies the collapse of a given region. 
As a result $\delta_{\rm c}$ depends on mass and this changes the mass function \citep{DelPopolo1999,DelPopolo2000}.

Similarly, \cite{Sheth2001} showed that moving from a spherical to an elliptical geometry, the collapse depends on the 
initial overdensity and shear, and $\delta_{\rm c}$ depends on mass. The mass function obtained with the elliptical 
collapse was shown to be in agreement with N-body simulations \citep{Sheth1999} (ST). 
However, a deeper analysis of those semi-analytic models for the mass function showed some problems: the PS MF, as 
already reported, overpredicts the MF at all high and medium masses \citep{Reed2003}, and even the ST MF 
overpredicts the halo number at large masses \citep{Lukic2007}. The situation worsens if one studies the PS and ST MF 
evolution. 
Simulations have been used to obtain a better understanding of the MF at low redshifts and of its evolution at high 
redshifts. \citep{Jenkins2001} tested the ST MF up to $z=5$ and down to $\simeq 3 \times 10^{11} M_{\odot}$, 
finding discrepancies only at uncommon (rare) density enhancements. \cite{Reed2003,Reed2007} found that the PS MF 
underestimates the rarest haloes in their simulations by a factor of $\simeq 5$, namely proto-galaxies at $z=30$, and 
galaxies with masses $\simeq 10^{11} h^{-1} M_{\odot}$ at $z=10$. Haloes hosting star populations at $z=30$ were 
underestimated by a factor of $\simeq 2$. The ST MF had a much better performance than the PS MF, but its predictive 
power decreased with increasing masses and redshifts. Their overestimation goes up to a factor of $\simeq 3$ for the 
rarest haloes \citep{Reed2007,Lukic2007}. 
In the case of the Bolshoi simulation \citep{Klypin2011}, the discrepancy is smaller than 10\% at $z=0$ in the 
mass range $5\times 10^9 - 5 \times 10^{14}~M_{\odot}$, while at $z=10$ the ST MF gives $\simeq 10$ 
times more haloes than simulations. \cite{Watson2013} compared several MF 
\citep{Press1974,Sheth2001,Reed2003,Warren2006,Reed2007,Crocce2010,Bhattacharya2011,Courtin2011,Angulo2012} 
with their Friends-of-Friends (FoF) MF (see the following) at redshift $z=0$. 
Excluding the PS MF, the remaining agree to $\simeq 10\%$ for masses $\leq 10^{15} M_{\odot}$.

The examples above show how an imprecise MF produces incorrect predictions, especially for halo numbers at high 
redshift (e.g., the number density of high redshift QSOs) or for astrophysical phenomena happening at high redshift 
(reionization scenarios and reionization history).

The universality of the MF, that is its independence on cosmology and redshift, poses an important issue studied by 
several authors \citep[e.g.][]{Reed2007,Tinker2008,Crocce2010,Bhattacharya2011,Courtin2011,Watson2013}. 
Obviously a universal MF would avoid the need to use N-body simulations to study it for different cosmologies and for 
its time evolution.

{
The majority of numerical simulations identify halos by one of two different techniques: either
friends-of-friends (FOF) or spherical overdensity (SO) algorithms. The FOF method identifies halos
by a percolation technique, connecting particles, within a certain distance (the linking length b) to each other, in the same halo. The linking length is typically chosen between $b = 0.15$ and $b = 0.2$, where $b$ is defined with respect to the mean interparticle spacing.
The FOF halo mass function scales very close to the ”universal” behaviour \cite{Klypin2011}.

The SO method first finds the halo centre from potential minimum or most
bound particle to identify haloes with spheres reaching a threshold density, given
with respect to either critical or background density. Typical ratios with respect to the critical density $\rho_{\rm c}$ are chosen from $\Delta=$ 200 to 500
(or higher for clusters). SO-based halo mass
functions are not universal especially at higher $z$.

The resulting two mass functions at $z = 0$ display close similarity for lower mass haloes, while the FOF case yields greater MF in higher-mass haloes \cite{Watson2013}.
At higher redshifts, the SO MF falls to $\sim 80 \%$ of the FOF at lower masses, with a stronger marked decrease at higher masses \cite{Watson2013}.

The disagreement between FOF and SO is likely related to the tendency of FOF to link structures before they become a part of a virialized halo, a feature more frequent with the rare most massive halos, that tend to be out of equilibrium and in the process of merging. As a result, FOF masses are artificially inflated.

However, for isolated, relaxed halos, SO and FOF masses are strongly correlated \cite{White2001,Tinker2008}. In cosmological simulations the mapping between the two halo
definitions can be considered one-to-one at the 5\% level of accuracy \cite{Lukic2009}. Nonetheless, a fair fraction
of halos in simulations are irregular: currently favoured cosmologies yield 15\%-20\% of FOF halos with irregular substructure with linking length $b = 0.2$ or two or more major halo components linked
together \cite{Lukic2009}. Such irregular halos not only fail to follow the simple SO to FOF mapping, but also lack a clear definition for halo
mass.
}

Most numerical simulations have used FoF masses with linking length $b=0.2$. 
As several authors showed \citep{Sheth1999,Jenkins2001,Reed2003,Warren2006}, the FoF MF obtained in cosmological 
simulations can be fitted by a function, $f(\sigma)$, of the variance of the linear density field $\sigma$. 
It has therefore been shown to be universal, independent of cosmology and redshift changes.

\cite{Jenkins2001} claimed a universal behaviour of the mass function within 20\%. \cite{White2002} showed the 
existence of deviations from universality and subsequent studies 
\citep[e.g.][]{Reed2007,Tinker2008,Crocce2010,Bhattacharya2011,Courtin2011} showed that the MF is not universal beyond 
the 5-10\% level.

Further studies on the FoF MF showed evidences of non-universality in the time evolution 
\citep{Crocce2010,Bhattacharya2011}, or when changing the cosmology \citep{Courtin2011}, while others showed 
an almost universal behaviour \citep{Watson2013}.
\footnote{According to \cite{Courtin2011}, the MF changes with cosmology because of the change of the collapse 
threshold $\delta_{\rm c}$. Cosmologies having smaller values of $\delta_{\rm c}$ than that of the SCDM allow 
structures forming earlier with respect to the SCDM model, giving rise to a different halo MF. 
In order to have an exact universality, one needs a mechanism able to eliminate the effect of the past evolution of 
structure formation on the MF \citep{Courtin2011}.}

\cite{Reed2007} found a violation of universality at high redshift, due to the effective spectral index $n_{\rm eff}$. 
Results of \cite{Lukic2007} were consistent with those of \cite{Reed2007}, but only at $z \leq 5$. 
\footnote{They studied the mass function in the mass range $10^7<M<10^{13.5}~h^{-1}~M_{\odot}$.} 
Their results exhibited a $\simeq 5\%$ residual in comparison to $z=0$, in agreement with the MF of \cite{Warren2006}. 
The FoF MF of the Millennium simulation increased by 20\% for the redshift range $z=0-10$ \citep{Reed2007} and, when 
corrected for "spurious FoF linking between haloes", showed the same evolution (20\%) in the range $z=0-1$ 
\citep{Fakhouri2008}.

Also \cite{Crocce2010} and \cite{Bhattacharya2011} found deviations from universality even for the FoF MF and 
provided redshift-dependent fits. 
\cite{Bhattacharya2011} proposed a redshift dependent FoF MF with an accuracy of $\simeq 2\%$ at $z=0$ and an 
evolution up to 10\% in the range $0<z<2$, $6 \times 10^{11} - 3 \times 10^{15}~M_{\odot}$. 
Similarly, \cite{Crocce2010} provided a fitting function that is accurate to 2\% in the ranges 
$0 \lesssim z \lesssim 1$ and $10^{10}\lesssim M \lesssim 10^{15.5} h^{-1} M_{\odot}$. 
\cite{Angulo2012} used the {\it Millennium-XXL} simulation to obtain a universal mass function accurate to 5\% in the 
mass range studied. \cite{Courtin2011} found a universal behaviour of the FoF MF in the $\Lambda$CDM model to a 5-10\% 
level. 
For different cosmologies they found deviations from the universal behaviour at 5\% level. 
\cite{Watson2013} calculated the MF from their simulations using two different halo-finding methods: the FoF and the 
Spherical Overdensity (SO). The FoF MF was found weakly dependent on redshift and was well represented by a universal 
fit. A universal FoF MF fits to 10\% the MF on their mass and redshift range. 
The MF universality {\it across redshift} was therefore valid only in some conditions and at a certain precision level. 
The SO MF had instead a redshift-dependent evolution. The proposed fit for the SO mass function was found valid within 
$\simeq 20\%$ in the redshift range $0 \lesssim z \lesssim 20$, while the universal function fit was good within 
$\simeq 10\%$.

Their result concerning the SO MF agreed qualitatively with \cite{Tinker2008}, who found clearer evidences for 
deviations from universality in the case of the SO haloes with respect to FoF haloes. 
In their simulation the redshift dependence comes from the parameter $\Omega_{\rm m}$. 
Their fit to the MF at $z=0$ in the mass range $10^{11}<M<10^{15}~h^{-1}~M_{\odot}$ was valid within 5\%.

According to \cite{Klypin2011}, the difference in behaviour, namely the almost universality of the FoF MF and 
non-universality of the SO MF, is related to the fact that FoF links structures before they merge to form a virialized 
halo. This produces an "inflation" of FoF masses, which, combined with the steep decrease of the MF, gives rise to an 
increase of the haloes number density.

In this paper, we will use the model of \cite{DelPopolo1998,DelPopolo1999,DelPopolo2000,DelPopolo2002b} and 
\cite{DelPopolo2006b,DelPopolo2006c} to show the evolution of the mass function that can be obtained in a modified 
spherical collapse model, taking into account the proto-structures' angular momentum acquired by tidal interaction with 
neighbouring objects and dynamical friction \citep[see also][]{Cardone2011a,Cardone2011b}.

In Section~\ref{sect:barrier} we find and discuss the barrier that is used in Section~\ref{sect:mf} to determine the 
multiplicity and the mass function. Section~\ref{sect:results} is devoted to the results, and 
Section~\ref{sect:discussion} to the discussion.

\section{ESF choices of barrier}\label{sect:barrier}
The semi-analytic model called extended Press-Schechter approach, based on stochastic processes, also known as 
"excursion set formalism" (ESF) \citep{Bond1991,Lacey1993} is often used to model the halo formation statistics and 
mergers.
\footnote{The term "excursion set formalism", introduced by \cite{Bond1991}, comes from the name of the regions, 
excursion sets, upon which the determination of the mass function is based, in the stochastic processes theory, and 
characterized by a linear density contrast $\delta_{\rm l}$ larger than the threshold $\delta_{\rm c}$.}

The halo statistics is obtained from the statistical properties of $\overline{\delta}(R_{\rm f})$, the average 
overdensity within a window of radius $R_{\rm f}$. Density perturbations are represented by a Gaussian density field, 
smoothed with a filter (e.g., top-hat or $k$-space filters, the latter being a top-hat filter in momentum space). 
In a hierarchical universe, $\overline{\delta}(R_{\rm f})$ vs $R_{\rm f}$ is a random walk 
\cite[see][]{DelPopolo2007a}; in the excursion set approach, a halo forms when the random walk crosses a 
threshold value, or barrier, $\delta_{\rm c}$. Other quantities are often used (e.g., the mass variance $S$) 
instead of the filtering radius.

In this framework, the first key quantity, the probability that a random walk first crosses the threshold (barrier) 
between $S$ and $S+dS$, is called the first-crossing distribution. The last key quantity, known as multiplicity 
function, is related (as we will see in the following) to the mass function.

The Press-Schechter MF is re-obtained in the ESF by studying random walks and flat barriers. In order to improve the PS 
formalism, random walks with a non-flat barrier were considered. They are usually dubbed {\it moving} barrier as the 
barrier changes (moves) with $S$ or mass.

The first study on the subject was \cite{DelPopolo1998}. They found that the collapse threshold becomes mass 
dependent. In particular, they showed that in the set of objects that collapse at the same time, those less massive 
must have initially been denser than the more massive ones in order for the former to hold themselves together against 
stronger tidal forces. 
\cite{DelPopolo1999} and \cite{DelPopolo2000} showed the new threshold (barrier) to give rise to a MF in good 
agreement with observations.

The mass dependence in the barrier solves the problems of the PS MF, suppressing low mass haloes abundance and 
increasing it for massive ones (with respect to the PS MF) \citep{Zhang2008b}.

Such moving threshold is given by: 
\begin{align}\label{eqn:barrier}
 \delta_{\rm cm} & = \delta_{\rm c}(z) \left[ 1 + \int_{r_{\rm i}}^{r_{\rm ta}}
                     \frac{r_{\rm ta}L^2\cdot{\rm d}r}{G M^3 r^3}\right]\nonumber \\
                 & = \delta_{\rm c}(z) \left[ 1+ \frac{8G^2}{\Omega_{\rm m,0}^3 H_0^6 r_{\rm i}^{10} \overline{\delta} 
                     (1+\overline{\delta})^2} \int_{r_{\rm i}}^{r_{\rm ta}} \frac{L^2 \cdot {\rm d}a}{a^3}\right] \\
                 & = \delta _{\rm c}(z) \left(1+\frac{\beta}{\nu^\alpha}\right)\;, \nonumber
\end{align}
where $\delta_{\rm c}(0)\simeq1.686$ is the critical threshold for the spherical collapse in an Einstein-de Sitter 
(EdS) model, $r_{\rm ta}$ is the turn-around radius, $r_{\rm i}$ the initial radius, $L$ the angular momentum acquired 
by the proto-structure, $a$ the expansion parameter, $H_0$ the Hubble constant and $\Omega_{\rm m,0}$ the density 
parameter today. 
The values of the parameters are $\alpha=0.585$ and $\beta=0.46$ \citep[e.g.][]{DelPopolo2002b,DelPopolo2006c}. 
In Eq.~(\ref{eqn:barrier}), the quantity $\nu=\left(\frac{\delta_{\rm c}}{\sigma}\right)^2$ is used, where $\delta_{\rm c}$ 
is the critical overdensity needed for collapse in the spherical model and $\sigma(M)$ is the r.m.s. of density 
fluctuations on a comoving scale including a mass $M$.%
\footnote{Note that the usual definition of $\nu$ gives $\nu=\frac{\delta_{\rm c}}{\sigma}$ \cite[see][]{DelPopolo1996}. 
The definition used in the text follows the ST notation.}

In a later work, \cite{Sheth2001} used an ellipsoidal collapse, finding
\begin{equation}\label{eqn:barrier1}
 \delta_{\rm ec} = \delta_{\rm c}(z) 
                   \left[ 1+{\beta_1} \left(\frac{\sigma^2}{\delta_{\rm c}(z)^2} \right)^{\alpha_1} \right]
                 = \delta_{\rm c}(z) \left( 1+\frac{\beta_1}{\nu^{\alpha_1}} \right)\;,
\end{equation}
with $\alpha_1=0.615$ and $\beta_1=0.485$. 

To Eqs.~(\ref{eqn:barrier}) and (\ref{eqn:barrier1}) correspond the two barriers
\begin{equation}\label{eqn:barrierv}
 B(M)=\sqrt{a} \delta _{\rm c}(z) \left(1+\frac{\beta}{a \nu^\alpha} \right)\;,
\end{equation}
and
\begin{equation}\label{eqn:barrierv1}
 B(M)_{\rm ST}=\sqrt{a_1} \delta _{\rm c}(z) \left(1+\frac{\beta_1}{a_1 \nu^{\alpha_1}}\right)\;,
\end{equation}
which give rise to a mass function in good agreement with simulations, when accurate values of $a$ are 
chosen, as shown by \cite{DelPopolo2006c} and \cite{Sheth2001}.%
\footnote{\label{fn:values}In the case of Eq.~(\ref{eqn:barrierv}), $a=0.67$ and for Eq.~(\ref{eqn:barrierv1}), $a_1=0.707$.}

Changing the shape of the barrier (from flat to increasing with $S$) allows to incorporate several physical effects, 
from fragmentation and mergers \citep{Sheth2001}, to the effects of tidal torques \citep{DelPopolo1998}, 
cosmological constant \citep{DelPopolo2006c} and dynamical friction (see the next sections).

As Eqs.~(\ref{eqn:barrierv}) and (\ref{eqn:barrierv1}) show, the barrier found by \cite{Sheth2001}, 
Eq.~(\ref{eqn:barrierv1}) and \cite{DelPopolo1998,DelPopolo1999}, Eq.~(\ref{eqn:barrierv}), are almost identical, as 
discussed in \cite{Sheth2002}.

In the case of a $\Lambda$CDM model, the collapse threshold, calculated in \cite{DelPopolo2006c}, is given by
\begin{equation}\label{eqn:barrierl}
 \delta_{\rm cm1} = \left[1+\int_{r_{\rm i}}^{r_{\rm ta}} \frac{r_{\rm ta} L^2 \cdot {\rm d}r}{G M^3 r^3}
 + \frac{\Lambda r_{ta}^3}{6GM} \right]
 = \delta_{\rm c}(z) \left(1+\frac{\beta}{\nu^\alpha}+ \frac{\Omega_{\Lambda}\beta_2}{\nu^{\alpha_2}}\right)\;,
\end{equation}
where $\alpha_2=0.4$, $\beta_2=0.02$ and $\Omega_{\Lambda}$ is the contribution of the cosmological constant $\Lambda$ 
to the density parameter.

The barrier given by Eq.~(\ref{eqn:barrierl}), depending on the proto-structures' acquired angular momentum and on the 
cosmological constant, can be further improved by taking into account another important effect on the collapse of 
proto-structures: dynamical friction (hereafter DF). This was done in \cite{DelPopolo2006b}, and is given by
\begin{align}\label{eqn:barrierf}
 \delta_{\rm cm2} & = \delta_{\rm co}\left[
                      1+\int_{r_{\rm i}}^{r_{\rm ta}} \frac{r_{\rm ta} L^2 \cdot {\rm d}r}{G M^3 r^3} +\Lambda \frac{r_{\rm ta} r^2}{6 G M}+ 
                      \frac{\lambda_{o}}{1-\mu(\delta)}\right] \nonumber \\
                  &   \simeq \delta_{\rm co} \left[
                      1+\frac{\beta}{\nu^{\alpha}}+\frac{\Omega_{\Lambda}\beta_2}{\nu^{\alpha_2}} + 
                      \frac{\beta_3}{\nu^{\alpha_3}}\right]\;,
\end{align}
where $\mu(\delta)$ is given in Eq.~(29) of \cite{Colafrancesco1995} and Appendix~\ref{sect:angmom}, and 
$\lambda_{o}=\epsilon_o T_{co}$, $\epsilon_o$ being proportional to the DF coefficient $\eta$ (see 
Appendix~\ref{sect:angmom} and Eq.~(23) of \cite{AntonuccioDelogu1994}) and $T_{co}$ being the DF-less perturbation 
collapse time (see Appendix~\ref{sect:angmom} and Eq.~(24) of \cite{AntonuccioDelogu1994}). The angular momentum $L$ is 
calculated as shown in \cite{DelPopolo2006b,DelPopolo2009,DelPopolo2009a} and in Appendix~\ref{sect:angmom}, while the 
DF term is obtained in \cite{AntonuccioDelogu1994} (see also Appendix~\ref{sect:angmom}).

\begin{figure}[tbp]
 \centering
 \includegraphics[scale=0.3]{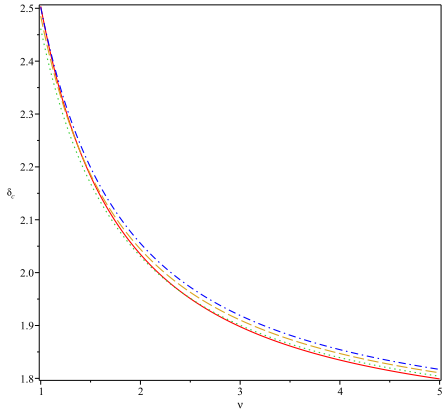}
 \par
 \caption[]{The collapse threshold $\delta_{\rm c}(\nu)$ as a function of $\nu$. The red solid line shows the result in 
 \cite{Sheth2001}, the green short-dashed line the results of \cite{DelPopolo1998}, taking into account the 
 effect of the tidal field, the orange long-dashed the result of \cite{DelPopolo2006c} taking into account the 
 effect of the tidal field and the cosmological constant, while the blue dot-dashed line takes into account the 
 effect of the tidal field, the cosmological constant and the DF.}
 \label{fig:deltac}
\end{figure}

Figure~(\ref{fig:deltac}) compares $\delta_{\rm c}(\nu)$ obtained by \cite{Sheth2001} by means of an ellipsoidal collapse 
model with the modified collapse thresholds obtained by Del Popolo. 
In the plot, the solid line represents the result of \cite{Sheth2001}, Eq.~(\ref{eqn:barrier1}), the dotted line 
Eq.~(\ref{eqn:barrier}) obtained by \cite{DelPopolo1998}, the dashed line Eq.~(\ref{eqn:barrierl}), namely the 
improvement of Eq.~(\ref{eqn:barrier1}) including $\Lambda$, and the dot-dashed line Eq.~(\ref{eqn:barrierf}), the 
improvement of Eq.~(\ref{eqn:barrierl}) accounting for the effect of DF.

All $\delta_{\rm c}(\nu)$ are monotonically decreasing functions of $\nu$ and mass $M$, and monotonically increasing 
functions of $S$, in contrast with other models \cite[e.g.][]{Monaco1997a,Monaco1997b}. Their behaviour tends to the 
typical value of the spherical collapse ($\delta_{\rm c}\simeq1.686$) for large $\nu$. This implies that less massive 
"peaks" (in the initial random field) form structures when crossing a higher threshold $\delta_{\rm c}$ than more 
massive ones.

Conversely, as high peaks are more probable in denser regions, the structure forming threshold $\delta_{\rm c}(M, z)$ 
is lower in overdense than underdense regions.

This reflects the different aspects of tides depending on the structures' mass. A peak acquires specific angular 
momentum $j$ proportionally to its turn-around time $t_{\rm ta}$, itself anti-correlated with the peak height: 
$j \propto t_{\rm ta} \propto \overline{\delta}(r, \nu)^{-3/2} \propto \nu^{-3/2}$. Hence smaller $\nu$ peaks are more 
sensitive to external tides, thus fixed time collapse leads them to be more overdense 
\citep{Hoffman1986,Ryden1988,Peebles1990,Audit1997,DelPopolo1998,DelPopolo2009,DelPopolo2010,DelPopolo2011}. 
Low-$\nu$ peaks tend to resist more to gravitational collapse than high-$\nu$ peaks because they acquire larger $j$, and 
thus need higher density contrast to collapse and form structures. This is why structures need, on average, higher 
density contrast to collapse for small scales, where shear is more important.

Those results agree with \cite{Peebles1990,Audit1997,DelPopolo2001,DelPopolo2002}. 
The latter found out that larger shear and angular momentum slow down the collapse. As smaller scales statistically 
hold larger shear and angular momentum, collapse of structures at those scales require a higher density contrast. Those 
results have been extended more recently by \cite{DelPopolo2013a,DelPopolo2013b,Pace2014b,DelPopolo2014b,Reischke2016a,Reischke2016b,Pace2017} for models with 
dark matter and dark energy.

Angular momentum possesses similar effects as a non-zero cosmological constant as, in particular, it especially slows down 
large mass structures' collapse, however the effect of the cosmological constant vanishes at high redshift 
(see also section~\ref{sect:results}). DF adds similarly to the angular momentum and the cosmological constant effects, 
cumulating into the moving barrier behaviour that reduces small haloes' abundance, and increases that of massive 
haloes, with respect to a flat barrier (PS mass function). A similar behaviour characterises the ST model of 
ellipsoidal collapse, caused in this case by the larger ellipticity carried by smaller haloes, leading to a larger 
collapse time \cite[see][]{Zhang2008a}.

\section{Multiplicity and mass function}\label{sect:mf}
In the ESF, the unconditional mass function $n(M,z)$, defined as the average comoving number density of haloes in a 
mass range $M - M+dM$ is \citep{Bond1991}
\begin{equation}\label{eqn:universal}
 n(M,z)=\frac{\overline{\rho}}{M^{2}}\left|\frac{d\log{\nu }}{d\log M}\right|\nu f(\nu)\;,
\end{equation}
where $\overline{\rho}$ is the background density. 
The quantity $f(\nu)$ is the so-called "multiplicity function", the distribution of the first crossing.

In the case of constant and linear barriers \citep{Sheth1998,Sheth2002}, one can obtain an analytical approximation for 
the first crossing. In other cases, the first crossing distribution can be obtained by generating a large ensemble of 
random walks. 
As shown by \cite{Sheth2002}, for a large range of moving barriers, one can approximate the first crossing distribution 
using the formula
\begin{equation}\label{eqn:distrib}
 f(S)dS=|T(S)|\exp{\left(-\frac{B(S)^{2}}{2S}\right)}\frac{dS/S}{\sqrt{2\pi S}}\;,
\end{equation}
where $T(S)$ can be obtained using a Taylor expansion of $B(S)$:
\begin{equation}\label{eqn:expans}
 T(S)=\sum_{n=0}^{5}\frac{(-S)^{n}}{n!}\frac{\partial ^{n}B(S)}{\partial S^{n}}
\end{equation}
where $S \equiv S_{\ast} (\frac{\sigma}{\sigma_{\ast}})^2 = \frac{S_{\ast}}{\nu}$, being 
$\sigma_{\ast}=\sqrt{S_{\ast}}$.
The multiplicity function is now given by $\nu f(\nu)=Sf(S,t)$.

Applying the previous methods to the barrier given by Eq.~(\ref{eqn:barrierv1}), the use of Eqs.~(\ref{eqn:distrib}) 
and (\ref{eqn:expans}) gives, at fifth order $n=5$,
\begin{equation}\label{eqn:sstt}
 \nu f(\nu)=\sqrt{a_1 \nu / 2 \pi}[1+\beta_1(a_1 {\nu})^{-\alpha_1} g(\alpha_1)] 
 \exp{\left(-0.5a_1\nu[1+\beta_1(a_1\nu)^{-\alpha_1}]^2\right)}\;,
\end{equation}
where
\begin{equation}
 g(\alpha)= \mid 1-\alpha +\frac{\alpha (\alpha -1)}{2!}- \ldots -\frac{\alpha(\alpha-1) \cdots (\alpha-4)}{5!} \mid\;,
\end{equation}
and the values of $a_1$, $\alpha_1$ and $\beta_1$ have been given previously. Then \cite{Sheth2002} give
\begin{equation}\label{eqn:sstt1}
 \nu f(\nu)\simeq \left(1+\frac{0.094}{\left(a\nu\right)^{0.6}}\right) 
                  \sqrt{\frac{a\nu}{2\pi}}
                  \exp{\left\{-\frac{1}{2}a\nu\left[1+\frac{0.5}{\left( a\nu \right)^{0.6}}\right]^{2}\right\}}\;,
\end{equation}
which is a good approximation to the first crossing distribution of the ellipsoidal barrier obtained through the 
simulations of unconstrained, independent random walks and that fitting the GIF simulations 
\cite[see figure~2 of][]{Sheth2002}
\begin{equation}\label{eqn:sstt2}
 \nu f(\nu)\simeq A \left( 1+\frac{1}{\left( a\nu \right)^{0.3}}\right)\sqrt{\frac{2 a\nu }{\pi }}\exp{(-a\nu/2)}\;,
\end{equation}
where $A=0.322$.

The same method, for the barrier taking into account the cosmological constant, given by Eq.~(\ref{eqn:barrierl}), 
gives
\begin{align}
 \nu f(\nu) & = A_1 \left(1+\frac{\beta g(\alpha)}{\left(a\nu\right)^{\alpha}}
               +\frac{\beta_2 g(\alpha_2)}{\left( a\nu \right)^{\alpha_2}}\right)
                \sqrt{\frac{a\nu }{2\pi }}\exp{\left\{-\frac{1}{2}a\nu\left[1+\frac{\beta}{\left(a\nu\right)^{\alpha}}
               +\frac{\beta_2}{\left( a\nu \right)^{\alpha_2}}\right]^{2}\right\}} \nonumber\\
            & \simeq A_1 \left(1+\frac{0.1218}{\left( a\nu \right)^{0.585}}
                    +\frac{0.0079}{\left( a\nu \right)^{0.4}}\right)
                     \sqrt{\frac{a\nu}{2\pi}}\exp{\left\{-0.4019 a \nu 
                     \left[1+\frac{0.5526}{\left(a\nu \right)^{0.585}}+
                     \frac{0.02}{\left( a\nu \right)^{0.4}}\right]^{2}\right\}}\;,
 \label{eqn:mia1}
\end{align}
The normalization factor $A_1=0.974$ must satisfy the constraint
\begin{equation}
 \int_0^{\infty} f(\nu) d\nu=1\;.\label{normmm}
\end{equation}

Finally, in the case of the barrier {
including, as in previous papers, }the effects of angular momentum and $\Lambda$, and {
now introducing the new effect of Dynamical Friction}, it is given by
\begin{align}\label{eqn:multiplic}
 \nu f(\nu) & \simeq A_2 \sqrt{\frac{a\nu}{2\pi}} \left(1+\frac{0.1218}{\left(a\nu\right)^{0.585}}
             +\frac{0.0079}{\left(a\nu\right)^{0.4}}+\frac{0.1}{\left(a\nu\right)^{0.45}}\right) 
              \nonumber\\
            & \quad\times 
               \exp{\left\{-0.4019 a \nu^{2.12}
               \left[1+\frac{0.5526}{\left( a\nu \right)^{0.585}}
              +\frac{0.02}{\left(a\nu\right)^{0.4}}+\frac{0.07}{\left(a\nu\right)^{0.45}}\right]^{2}\right\}}\;,
\end{align}
where $A_2=0.9
3702$ and {
$a$=0.707}. {
Thus Eq.~\ref{eqn:multiplic} presents our  mass function, calculated in Appendix~\ref{sec:MF}, whose validity shall now be confronted with simulations 
in the next section.}

In the following we will use the CDM spectrum of \cite[][Eq.~(G3)]{Bardeen1986}.

{
At this point, it is important to stress that all numerical constants (except $a$) derive from barrier calculations: condition (\ref{normmm}) gives the normalization constant $A$. Although the parameter 
$a$, which gives the number of high mass haloes, could also be obtained by the excursion set theory with a diffusing barrier, as shown by \cite{Maggiore2010c}, it was determined as a fit to the massive haloes number in the simulations of \cite{Sheth2001}, and thus could be interpreted as depending on the halo finder, which, in our case, similarly to  \cite{Bhattacharya2011}, is a FoF with a linking length of $b=0.2$.
}

{
Since the number of halos, defined via FoF or SO techniques, is more affected by the change of technique at low masses rather than at high 
masses (provided the mass resolution of the simulation is good enough), the value of $a$ is largely unaffected by the 
choice of the specific algorithm. Different considerations hold instead for the dependence of $a$ as a function of the 
linking length parameter $b$. Given a background cosmological model, the virial overdensity $\Delta_{\rm vir}$ can be 
evaluated, under certain assumptions, within the formalism of the spherical collapse model. Having this value and 
assuming for simplicity an isothermal profile for the dark matter halos, the mean density of the halo at the virial 
radius is $\rho_{R_{\rm vir}}=\bar{\rho}\Delta_{\rm vir}/3$. Assuming that the density at the virial radius is represented by that of 
two particles in a sphere of radius $b$, one can relate it to a mean separation between particles. For an EdS model, 
$b=0.2$, while for a $\Lambda$CDM cosmology with $\Omega_{\rm m}=0.25$ and $\Omega_{\Lambda}=0.75$, $b=0.156$. While the previously presented values proceed from the theoretical derivation of the linking length, in practice  the value of an EdS model is also usually used for 
$\Lambda$CDM cosmologies. As noted by \cite{Courtin2011}, deviations from universality for the FoF mass function can be 
minimised by using the appropriate value of $\Delta_{\rm vir}$, hence a correct value for $b$. Ref.~\cite{Courtin2011} 
also quantified this and showed that the best choice for $b$ (called $b_{\rm univ}$) can be obtained with the following 
relation:
\begin{equation}
 \left(\frac{b_{\rm univ}}{0.2}\right)^{-3} = 0.24\left(\frac{\Delta_{\rm vir}}{178}\right)+0.68\;.
\end{equation}
This empirical correlation can be explained in the light of the results of \cite{More2011}. The authors showed that at 
$z=0$, the FoF overdensity for $b=0.2$ is significantly larger than 178 and that it depends on $b$ and on the halo 
concentration. Their analysis led to the following expression relating $b$ to $\Delta$ (the enclosed FoF overdensity)
\begin{equation}
 \left(\frac{b}{0.2}\right)^{-3} = \frac{\Delta+1}{244.86}\psi(c_{\Delta})\;,
\end{equation}
where
\begin{equation}
 \psi(c) = \frac{c^2}{\mu(c)(1+c)^2}\;,
\end{equation}
and
\begin{equation}
 \mu(x) = \ln{(1+x)}-\frac{x}{1+x}\;.
\end{equation}
It is clear that the way the concentration parameter is defined will also affect the definition of the linking length. 
This implies therefore a dependence of the mass function on the linking length parameter $b$.

In the light of these considerations, it is reasonable to assume that using the appropriate linking length for a given 
cosmology, will leave $a$ unaffected. This could probably not be the case if, for a fixed cosmology, values of $b$ much 
different from the optimal one are used, due to the bridging problem affecting FoF methods. As also explained by 
\cite{More2011}, this issue needs to be investigated more deeply and quantitatively.
}

\section{Results}\label{sect:results}
In the present section, we will compare our mass function at $z=0$ with several mass functions obtained from 
simulations over a large mass range. We will then compare its evolution with the result of \cite{Klypin2011}, in 
the redshift range $0<z<10$ {
but over a more restricted mass range. Our choice was dictated by the lack of other exploitable simulations showing $z$-dependence: \cite{Reed2003} is old, and \cite{Watson2013} only plots some residuals and not the MF.}

\begin{figure}[htbp]
 \centering
 \includegraphics[scale=1]{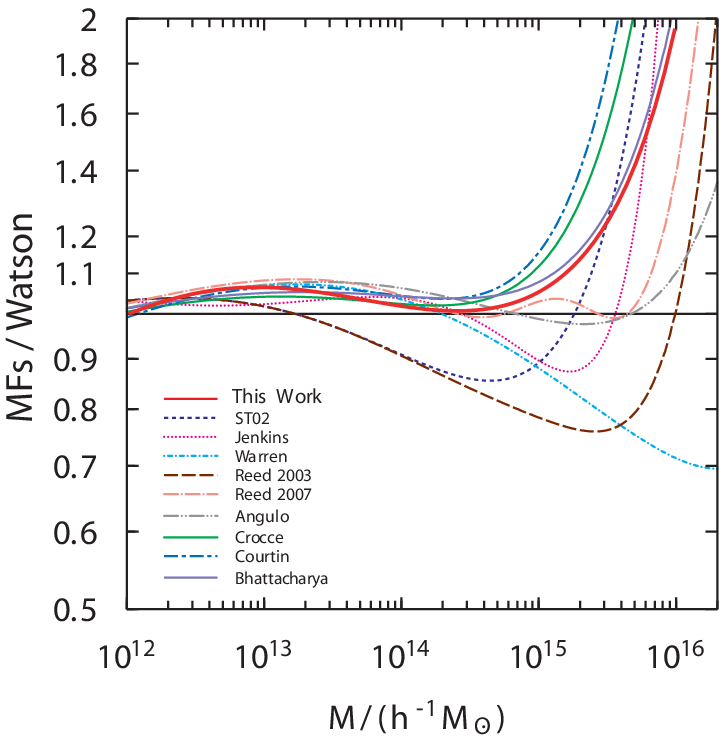}
 \includegraphics[scale=0.4]{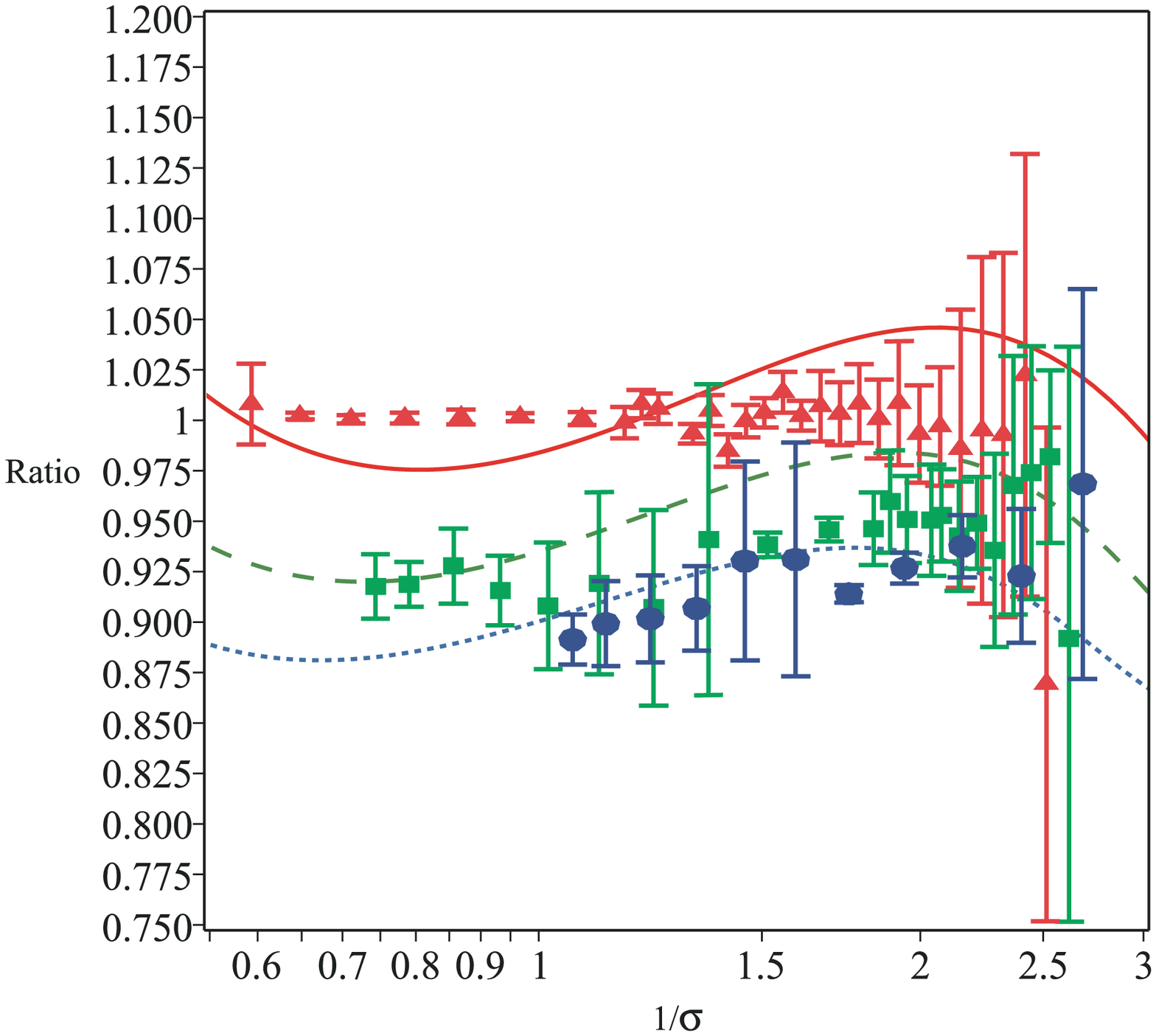}
 \par
 \caption[]{Left panel: ratios between our mass function and others from literature, and the Watson MF at $z=0$ 
 \citep{Watson2013}. 
  The red solid line represents the mass function proposed in this work; the blue short-dashed line the Sheth-Tormen 
 mass function \citep{Sheth2002}; the purple dotted curve the Jenkins mass function \citep{Jenkins2001}; the cyan 
 short-dashed-dotted curve the Warren mass function \citep{Warren2006}; the brown dashed and the pink dashed-dotted 
 curves the two Reed mass functions \citep{Reed2003,Reed2007}, respectively; the gray dashed-dot-dotted curve the 
 Angulo mass function \citep{Angulo2012}; the green and the violet solid curves show the Crocce \citep{Crocce2010} and 
 the Bhattacharya \citep{Bhattacharya2011} mass functions, respectively. Finally the light blue dashed-short-dashed 
 curve represents the Courtin mass function \citep{Courtin2011}. 
 Right panel: {
 comparison between our proposed mass function and that of Bhattacharya \cite{Bhattacharya2011} for 
 different redshifts. We show the ratio of our mass function at a given redshift with respect to the one 
 proposed by Bhattacharya, at $z=0$ for ensuring legible separation of the curves, in analogy to their Fig.~5. The red solid, green dashed and dotted blue lines show 
 the ratio at $z=0$, $z=1$ and $z=2$, respectively. The corresponding data set points are those obtained in 
 Bhattacharya's paper at $z=0$, $z=1$ and $z=2$, respectively.}
 }
 \label{fig:mf}
 \label{fig:ratioz}
\end{figure}

In the left panel of 
figure~\ref{fig:mf}, we plot the ratios of the mass function proposed by 
\cite{Jenkins2001,Sheth2002,Reed2003,Warren2006,Reed2007,Crocce2010,Bhattacharya2011,Courtin2011,Angulo2012} with 
\cite{Watson2013}, valid in the range $-0.55< \log{\sigma^{-1}} <1.31$ which at $z=0$ 
corresponds to the mass range $1.8 \times 10^{12}-7.0 \times 10^{15}~h^{-1}~M_{\odot}$. 
\cite{Sheth2001,Jenkins2001,Sheth2002,Warren2006,Courtin2011} give a universal mass function, while 
\cite{Reed2007,Crocce2010,Bhattacharya2011} provide an expression with z-dependent coefficients.%
\footnote{The z-dependence in \cite{Sheth2001,Jenkins2001,Sheth2002,Warren2006,Courtin2011} is parametrised through 
$\delta_{\rm c}(z)$.}
Apart from older mass functions, like that of \cite{Reed2003}, our mass function agrees {
on average} to a $\simeq 
{3}\%$ level 
in the mass range $10^{12}-10^{15}~h^{-1}~M_{\odot}$ with all the other mass functions. For 
$M> 10^{15}~h^{-1}~M_{\odot}$ our MF agrees well with that of \cite{Bhattacharya2011} (see the following). 
The latter MF also agrees with simulations data of more recent mass functions to better than 2\% accuracy, while 
\cite{Angulo2012}, based on the Millennium-XXL simulations, agrees to 5\% with those simulations, \cite{Courtin2011} 
do so to $5 - 10\%$, and \cite{Crocce2010} to 2\%. 
In the right panel of 
{
figure~\ref{fig:ratioz}}, we plot the ratio between the Bhattacharya MF and ours. {
The discrepancy remains under 3\% for the whole mass range $10^{12}\textnormal{---}10^{16}~h^{-1}~M_{\odot}$}.

\begin{figure}[tbp]
 \centering
 \includegraphics[scale=0.35]{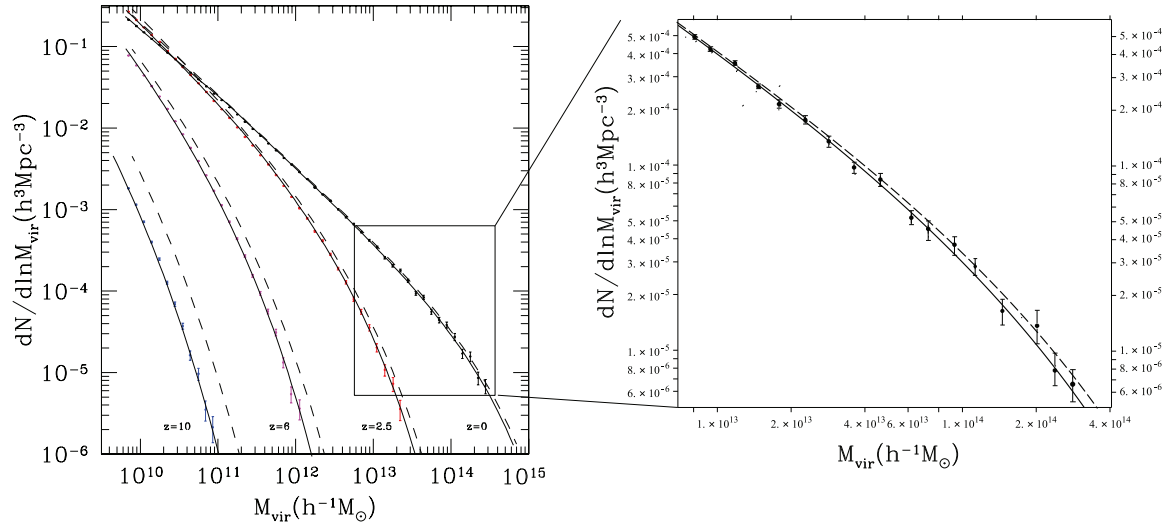}
 \par
 \caption[]{Comparison of the mass functions of \cite{Sheth2002} and this work with the Bolshoi MF. Left panel: 
 the dashed line represents the ST MF, while the solid line ours. Diamonds represent the Bolshoi MF. From left to right 
 $z$ receives the values 10, 6, 2.5, 0. {
 Right panel:  zoom of the $z=0$ MF.}}
 \label{fig:mf_sim}
\end{figure}

\begin{figure}
 \centering
 \includegraphics[scale=0.4]{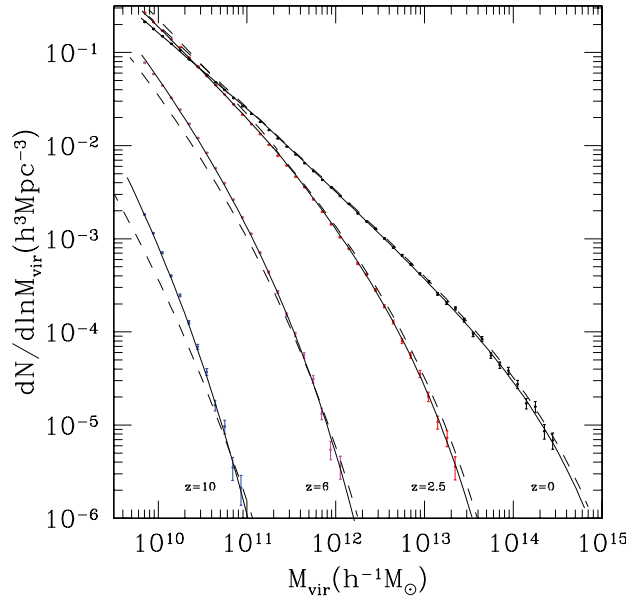}
 \par
 \caption[]{{
 Comparison of our analytic mass function with the Bolshoi's of \cite{Klypin2011}. Curves and symbols are as in the left panel of figure~\ref{fig:mf_sim}, except that now the 
  dashed line represents the correction by \cite{Klypin2011} to the ST MF.} 
}
 \label{fig:mf_sim1}
\end{figure}

In the panels of figure~\ref{fig:mf_sim} {
and \ref{fig:mf_sim1}}, we plot the MF for different redshifts. The solid lines show our MF related 
to the multiplicity function given by Eq.~(\ref{eqn:multiplic}), while on the left panel, the dashed line displays the 
MF related to the multiplicity function (Eq.~(\ref{eqn:sstt2})) of \cite{Sheth2002} and the diamonds with error-bars 
represent the MF obtained in the Bolshoi simulation by \cite{Klypin2011}.\footnote{ 
Note that the use of \cite{Sheth2002}, now considered not particularly accurate, is shown for comparison with the work of \cite{Klypin2011} only, as they introduce a correction that performs more poorly than our MF, as seen below.} 
The MF redshift dependence comes from $\delta_{\rm c}(z)$. At $z=0$, the rightmost dashed curve, the mass function 
of \cite{Sheth2002}, deviates from simulation data by less than 10\% in the mass range 
$M_{\rm vir} = 5 \times 10^9 - 5 \times 10^{14}~h^{-1}~M_{\odot}$. 
At higher redshifts, \cite{Sheth2002} overpredicts simulation results, and that overprediction increases with redshift. 
At $z=6$, and for masses $M_{\rm vir} = 1 - 10 \times 10^{11}~h^{-1}~M_{\odot}$, \cite{Sheth2002} predicts 1.5 more 
halos than the simulation. The situation is much worse at $z=10$ since the MF of \cite{Sheth2002} predicts 10 times 
more haloes than the simulation. Our mass function (solid line) is in good agreement with the simulation with 
deviations  {
$\lesssim 3\%$, calculated by average on each curve from the central values, intersecting all the error bars}. 

{

In addition, as already reported, our result is also in agreement with simulations by \cite{Tinker2008} and 
\cite{Cohn2008}. In contrast, our result shows an increasing overprediction of the mass function of 
\cite{Sheth2002} going to larger masses and higher redshifts, together with a steepening of the MF with mass (there is a greater disagreement at larger masses).
As stressed by \cite{Tinker2008}, this behaviour reveals MF non-universality (i.e. dependence on redshift and cosmology), as the MF dependence with $z$ doesn't follow just the amplitude of $\sigma(M)$ (see definition in Appendix \ref{sec:MF} or \cite{Klypin2011}'s Appendix B). 
Although our MF (Eq. \ref{eqn:multiplic}) displays a universal-like structure, similarly to ST \cite{Sheth2002}, its dependence on cosmology and $z$ differs from universal. This can be understood as follows.

The universality of the Sheth--Tormen (ST) formula \cite{Sheth1999,Sheth2001} proceeds from the multiplicity function only depending on the collapse threshold $\delta_{\rm c}$, a function both of $z$ and the cosmology. \cite{Courtin2011}  stressed that the literature often neglects the cosmology dependence encoded in $\delta_{\rm c}$, because the Standard Cold Dark Matter (SCDM) scenario spherical collapse model, predicting $\delta_{\rm c} = 1.686$ and
$\Delta_{\rm vir} = 178$ constant in redshift, has long been the structure formation studies' reference cosmology, and this $\delta_{\rm c}$ value has been long kept. It has also been shown that taking into account the cosmology dependence of 
$\delta_{\rm c}$ in non-standard cosmology provides a good agreement between analytic and numerical MF 
\citep{Pace2014}. 
Alternatively, the mass function measured in numerical simulations has been directly fitted by multiplicity functions depending only on $\sigma$ \cite{Jenkins2001,Linder2003a,Warren2006}, the resulting MF being thus manifestly independent of  cosmological and redshift evolution, what is commonly understood as "universality" \citep[cosmology and redshift independence of the relation between the linear and non-linear growth of structures;][]{Francis2009a}.

\cite{Courtin2011} showed a direct effect of evolution of $\delta_{\rm c}$, and $\Delta_{\rm vir}$ with $z$ and cosmology on the MF, as haloes virialization, which depends upon cosmology and redshift, plays a role in determining it. These results demonstrate the importance of nonlinear effects in the MF prescription. 

To summarize, the cosmology dependence of the non-linear collapse and virialization process creates deviations from universality, from cosmological model dependence of the spherical collapse threshold, responsible for deviations in the high-mass end of the MF, and accounting for it
reduces MF discrepancy between models. 

}

Ref.~\cite{Klypin2011}, preceded by several authors \citep[e.g.][]{Reed2003}, proposed an improvement to the ST mass 
function, multiplying it by a correction factor
\begin{equation}\label{eq:corrfunc}
 F(\delta) = \frac{(5.501 \delta)^4}{1+(5.500 \delta)^4}\;,
\end{equation}
where $\delta$ is the linear growth factor normalized to unity today. \cite{Klypin2011} claimed that the corrected mass 
function deviates by less than 10\% from simulations in the mass range 
$5\times 10^{9} - 5\times 10^{14}~h^{-1}~M_{\odot}$. 
This is presented in {
figure~\ref{fig:mf_sim1}}, showing the comparison of the Bolshoi data 
(diamonds) with the correction by \cite{Klypin2011} (dashed line), and the result of our model (solid line). {
Note however that calculating the average error on $z=10$ yields 38\% for the correction by \cite{Klypin2011} instead of the claimed 10\%, as opposed to our 3\%.} 
It is evident there that our MF gives a much better result than the correction by \cite{Klypin2011}.

The results displayed in figures~\ref{fig:mf_sim} and \ref{fig:mf_sim1} show that the MF generated from our barrier is 
in good agreement both with simulations at $z=0$ and with their redshift evolution with a precision of the order of {
3\%}. By contrast, the left panel of figure~\ref{fig:mf_sim} shows the discrepancy between the predictions by 
\cite{Sheth2002} and simulations. \cite{Sheth1999} introduced the effects of asphericity considering an intuitive 
parametrization of an elliptical collapse without taking into account the interaction with neighbours (isolated 
spheroid), nor considering the effect of $\Lambda$ in the barrier. Thus, while their model is an improvement on the 
spherical collapse based PS model, it remains partial, leading to the aforementioned MF discrepancies with simulations. 
Our model, in contrast, takes into account the cosmological constant $\Lambda$, the effects of dynamical friction and 
of angular momentum, acquired through tidal torques. Such improvements give rise to a MF in good agreement with 
simulations, as we will detail in the following. 
Moreover, barriers increasing with $S$, like ours, allow mergers and fragmentation, whereas barriers decreasing with 
$S$ \cite[e.g.,][]{Monaco1997a,Monaco1997b}, are characterised by the facts that all walks cross them and 
fragmentation is not allowed.

{
In figure~\ref{fig:ratio} we show the fractional accuracy of our proposed mass function with respect to the 
numerical mass function of \cite{Klypin2011} at $z=0$. The points and the error-bars correspond to the values from 
\cite{Klypin2011}. Note how all the error-bars intersect our curve. Errors are relatively small for low-mass objects 
and gradually increase towards higher masses. This is easily explained taking into account that the higher the mass, 
the lower is the number of objects. The point most differing from our mass function has a mass 
$M\approx 1.75\times 10^{14}~h^{-1}M_{\odot}$. Over all the points, the accuracy of the mass function presented in this 
work with respect to the simulation used as comparison is, on average, better than 4\%.

\begin{figure}
 \centering
 \includegraphics[scale=0.5]{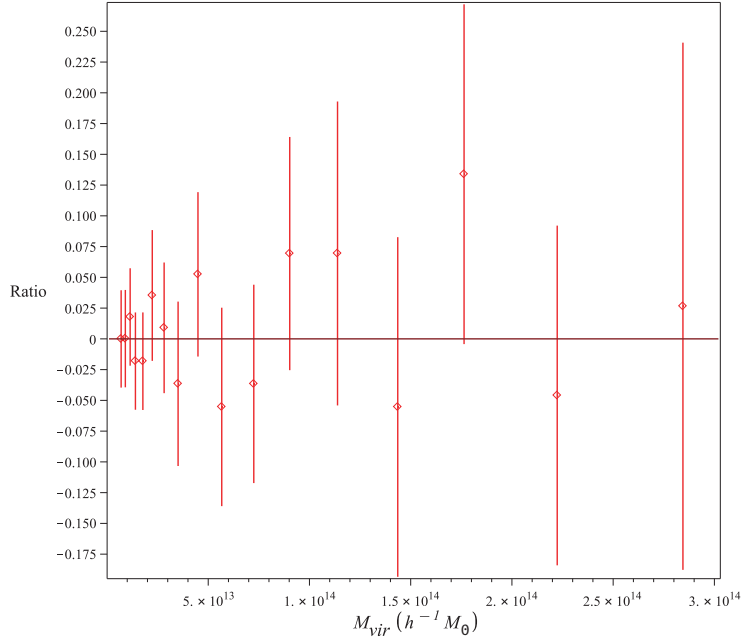}
 \caption{
 Fractional ratio between the numerical MF by \cite{Klypin2011} and our proposed mass function at 
 $z = 0$.}
 \label{fig:ratio}
\end{figure}

It is also interesting to evaluate how the ratio between of our proposed mass function and that of Bhattacharya changes 
in time. This is shown in figure~\ref{fig:ratioz}'s right panel where we consider the following redshifts: $z=0,1,2$. Points and 
corresponding error bars are those obtained in Fig.~5 of \cite{Bhattacharya2011}.

The data sets at $z=1$ and $z=2$ demonstrate that redshift evolution is important and must be taken into account; in 
addition, over the range of redshifts considered, the overall agreement is of the order of 3\%, except for few points 
at higher masses (low $\sigma$) where the agreement is at the 5\% level.
}

Thus, a precise MF requires a precise determination of the barrier, whose shape depends on the effects of dynamical 
friction, the cosmological constant and angular momentum.

\begin{figure}
 \centering
 \includegraphics[scale=0.6]{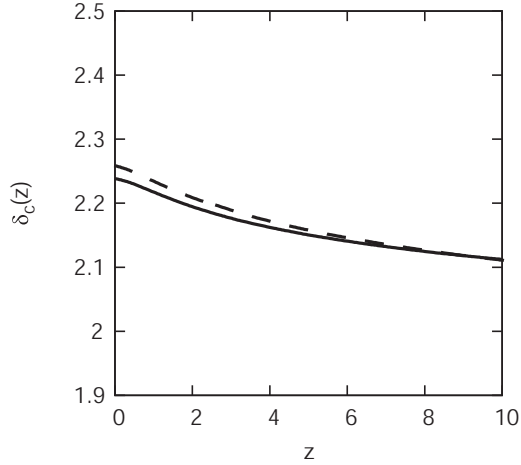}
 \par
 \caption[]{The collapse threshold in terms of $z$ when angular momentum and $\Lambda$ is taken into account (solid line) for a mass 
 $M=10^{11}~h^{-1}M_{\odot}$. The dashed line represents the same quantity when dynamical friction is taken into account.}
 \label{fig:dcL}
\end{figure}

{
At this stage it is important to emphasize the need for a new and precise MF, especially when it agrees with previous ones \citep[e.g., Bhattacharya's][]{Bhattacharya2011}.
Multiplicity functions presented in this paper, such as Bhattacharya's and except for our's (Eq.~\ref{eqn:multiplic}), are produced by high resolution N-body simulations fits, similar in functional forms to \citep[e.g. ][]{Sheth2002,Jenkins2001}. These fits have no theoretical foundations, revealing the importance of obtaining a realistic analytical form from {\it first principles}. Such form is both able to better "describe" simulations and physically motivated. The MF (\ref{eqn:multiplic}) obtained in this paper does provide an excellent prediction of high resolution simulations, and at the same time derives from solid physical and theoretical arguments.

On top of this theoretical advantage,  our approach can very accurately predict the dark matter halo distribution at much lower computational cost than high resolution simulations. {
This is because we can derive its functional form 
without having to rely on numerical results: it follows up directly by using an improved barrier.}

In conclusion, the excursion set approach, with a structure formation physics motivated barrier, produces an excellent approximation to the numerical multiplicity function: improving the barrier form (with more and more physical effects: angular momentum acquisition, non-zero cosmological constant, etc.) increases the approximation accuracy. Moreover, this method displays a remarkable versatility: any effect, such as the presence of a non-zero cosmological constant, is very easy to take into account by embedding it in the barrier.}

The role of angular momentum in shaping the MF was discussed in section~\ref{sect:barrier}, where we showed that it 
reduces or prevents structure formation, especially at small scales.

In addition to the $\delta_{\rm c}$ dependence on mass, one needs to take into account its time evolution. In 
figure~\ref{fig:dcL} we plot $\delta_{\rm c}(z)$, following \cite{DelPopolo2013a} for a $\Lambda$CDM model 
\cite[for a generalization to DE models, see][]{DelPopolo2013b,Pace2014b}, taking into account the angular momentum  and $\Lambda$ for 
a mass $M = 10^{11}~h^{-1}~M_{\odot}$. The dashed line adds the effect of dynamical friction.
Angular momentum causes $\delta_{\rm c}$ to be a monotonic decreasing function of $z$ while it already is the case for mass 
$M$): $\delta_{\rm c}(z)$ is larger than $\delta_{\rm co}$ at all values of $z$. Thus structure formation is 
"suppressed" at high $z$ by angular momentum.%
\footnote{A similar result was obtained in \cite[][(Fig. 3)]{DelPopolo2013a}.} 
This explains why our MF predicts a smaller abundance than that of \cite{Sheth2002} with increasing $z$.

The effect of the cosmological constant can be understood from Eq.~(\ref{eqn:coll}) (discussed in detail in 
Appendix~\ref{sect:angmom}). The term involving the cosmological constant has the same effect as those involving 
angular momentum and DF, namely, slowing down the collapse 
\citep{DelPopolo1997,DelPopolo1998,DelPopolo1999,DelPopolo2002
}. 
This gives rise to a delay in large-scale structure formation, reducing their abundance and steepening the MF. This 
would also produce a larger proportion of high-$z$ haloes, that would be smaller than the resolution of simulations 
\citep{Feyereisen2015}. At the same time the cosmological constant clearly decreases the number of halos in the 
high-mass tail, (Feyereisen private communication and paper in preparation).

As discussed by \cite{Bhattacharya2011}, since at high redshifts the effect of $\Lambda$ decreases, the non-universal 
evolution with redshift should be suppressed. However, despite 
the effect of $\Lambda$ reducing with $z$, the 
non-universal behaviour persists at high $z$  due to other factors \cite[see][]{Courtin2011}.

Dynamical friction also slows down the collapse, similarly to the cosmological constant. 
Of the three effects taken into account, angular momentum is the strongest in slowing down the collapse, followed by 
dynamical friction.\footnote{
For an alternative description of the effects of tidal shear and angular momentum for the $\Lambda$CDM and 
dark energy models, we refer to \cite{Reischke2016a,Reischke2016b,Pace2017}.}

Before concluding, we want to point out that the agreement between our MF and the Bolshoi simulation data could be further
improved assuming a slight redshift dependence of $A_2$ and $a$, as done in many of the papers cited in the 
Introduction.

\section{Discussion}\label{sect:discussion}
The determination of a high precision mass function is of fundamental importance. 
After the first improvements of the PS MF by \cite{DelPopolo1999} and \cite{Sheth1999,Sheth2001,Sheth2002}, 
further advances came from N-body simulations, used as tools to calibrate proposed fitting formulas 
\citep[e.g.,][]{Jenkins2001,Reed2003,Reed2007,Warren2006,Tinker2008,Crocce2010,Bhattacharya2011,Courtin2011,Watson2013} 
and more recently from a new diffusing barrier 
\citep{Maggiore2010a,Maggiore2010c,Maggiore2010d,deSimone2011a,deSimone2011b,Ma2011b}.

Although MF obtained through simulations yield good approximations in several cases, the simulations black box nature, 
in which many physical effects are taken into account, makes it difficult to disentangle the role of those mechanisms 
in the shaping of the MF. In our approach, we selected physical effects known to play an important role in the shaping 
of the MF and explained with them why the PS MF gives bad fits to the observed MF, and why the ST MF has problems in 
reproducing it at high $z$. At the same time, the approach leaves us with a semi-analytical form of the MF in 
very good agreement with simulations. In this sense our result is much more physical than that of simulations and gives 
a fit to their results.

The paper shows that the introduction of a moving barrier makes the collapse threshold mass-dependent, contrary to the 
standard spherical collapse model, but in parallel with extended models where shear, tidal fields and/or angular 
momentum are taken into account.

An interesting feature of a moving barrier is the possibility to introduce effects such as mergers, tidal torques, 
dynamical friction, and cosmological effects such as the cosmological constant (note that, as also pointed 
out by \cite{Murray2013b}, \cite{Sheth2001} fitted their mass function with an EdS model).

The effect of introducing the cosmological constant remains minor compared with other effects such as tidal fields and 
angular momentum, but both the cosmological constant and the angular momentum slow down the collapse.

The positive consequence of these aspects is to solve the PS approach problems, in particular to reduce (increase) the 
number of objects at low (high) mass \citep{DelPopolo2002a,DelPopolo2006c}. A similar result has been found for the 
ellipsoidal collapse \citep{Sheth2001}.

The barrier for the first crossing shapes the mass function and modify its functional form with respect to the simple 
PS formulation. The improved mass function yields results in very good agreement with N-body simulations 
{
\citep{Klypin2011,Bhattacharya2011}, within 3\% level at $z=0$ and}  during its time evolution (see Figure~\ref{fig:mf_sim}). 
The general behaviour of our proposed mass function is in agreement with other functional forms proposed in 
literature, such as fits to numerical simulations (see Figure~\ref{fig:mf}) and shares with them, albeit at a lower 
level, the same problem: an excess of structures with respect to numerical simulations predictions at high redshifts.

\acknowledgments
The authors thank an anonymous referee for the useful comments that helped improving the scientific content of this work.\\
FP is funded by an STFC post-doctoral fellowship.
The work of M.Le~D. has been supported by PNPD/CAPES20132029. M.Le~D. also wishes to acknowledge IFT/UNESP.
\bibliographystyle{JHEP}
\bibliography{mf.bbl}

\appendix

\section{Inclusion of the angular momentum}\label{sect:angmom}

As already discussed, in hierarchical models, a perturbation collapses at a given time, when its overdensity exceeds a 
critical threshold. The barrier is the linear extrapolation of the threshold to the present time. We saw that 
\cite{DelPopolo1998} found a moving barrier, taking into account angular momentum acquisition. 
In \cite{DelPopolo2006b,DelPopolo2006c} the barrier was extended to account for the role of the cosmological constant 
and of DF.

The delay of collapse of a perturbation due to the acquisition of the angular momentum, the presence of dynamical 
friction, and a non-zero cosmological constant, can be obtained by solving the equation 
\citep{Kashlinsky1986,Kashlinsky1987,Lahav1991,Bartlett1993,AntonuccioDelogu1994,Peebles1993,DelPopolo2006b}:

\begin{equation}\label{eqn:coll}
 \frac{dv_r}{dt}=\frac{L^2(r,\nu )}{M^2r^3}-g(r) -\eta \frac{dr}{dt}+ \frac{\Lambda}{3}r\;,
\end{equation}
where $\Lambda$ is the cosmological constant, $g(r)$ the acceleration, $L(r,\nu)$ the angular momentum and $\eta$ the 
coefficient of dynamical friction. 
Recalling that the proper radius of a shell can be written as 
\begin{equation}
 r(r_{\rm i},t)=r_{\rm i} a(r_{\rm i},t)\;,
\end{equation}
where $a(r_{\rm i},t)$ is the expansion parameter and $r_{\rm i}$ the initial radius, and that the mass is given by 
\begin{equation}
 M=\frac{4\pi}{3}\overline{\rho}(r_{\rm i},t)a^3(r_{\rm i},t)r_{\rm i}^3\;,
\end{equation}
being the average density $\overline{\rho}(r_{\rm i},t) = \frac{\overline{\rho}_{\rm i}(r_{\rm i},t)}{a^3(r_{\rm i},t)} 
= \frac{\overline{\rho}_{\rm ci}(1+\overline{\delta})}{a^3(r_{\rm i},t)}$ and 
$\rho_{\rm ci}=\frac{3H_{\rm i}^2}{8\pi G}$, Eq.~(\ref{eqn:coll}) may be written as
\begin{equation}\label{eqn:sec}
 \frac{d^2a}{dt^2} = -\frac{H^2(1+\overline{\delta})}{2a^2}+\frac{4G^2L^2}{H^4(1+\overline{\delta})^2r_{\rm i}^{10}a^3} 
 - \eta\frac{da}{dt} + \frac{\Lambda}{3}a\;.
\end{equation}

For $\eta=0$ the two equations become
\begin{equation}\label{eqn:colls}
 \frac{dv_r}{dt} = \frac{L^2(r,\nu)}{M^2r^3} - g(r) + \frac{\Lambda}{3}r\;,
\end{equation}
and 
\begin{equation}\label{eqn:secs}
 \frac{d^2a}{dt^2} = -\frac{H^2(1+\overline{\delta})}{2a^2} + 
 \frac{4G^2L^2}{H^4(1+\overline{\delta})^2r_{\rm i}^{10}a^3} + \frac{\Lambda}{3}a\;.
\end{equation}

Integrating Eqs.~(\ref{eqn:colls}) and (\ref{eqn:secs}) we can write
\begin{equation}\label{eqn:collss}
 \left(\frac{dr}{dt}\right)^2 = \int \frac{2L^2(r,\nu)}{M^2r^3}dr - 2 g(r) + \frac{\Lambda}{3}r^2 - 2C\;,
\end{equation}
and
\begin{equation}\label{eqn:secss}
 \left(\frac{da}{dt}\right)^2 = \frac{H_{\rm i}^2(1+\overline{\delta})}{a} + 
 \int \frac{8G^2L^2}{H_{\rm i}^4(1+\overline{\delta})^2r_{\rm i}^{10}a^3} + \frac{\Lambda}{3}a^2 - 2C\;,
\end{equation}
where $C$ is the binding energy of the shell \citep[see][]{Bartlett1993} and can be obtained using the conditions 
$dr/dt=0$ for Eq.~(\ref{eqn:collss}) and $da/dt=0$ for Eq.~(\ref{eqn:secss}).

Integrating once more, recalling that $dr/dt=da/dt=0$, we get \citep{DelPopolo2002a}
\begin{equation}\label{eqn:temp}
 t_{\rm ta} = \int_{0}^{r_{\rm ta}}\frac{dr}{\sqrt{2\left[GM\left(\frac{1}{r}-\frac{1}{r_{\rm ta}}\right) + 
              \int_{r_{\rm ta}}^{r}\frac{L^{2}}{M^{2}r^{3}}dr + \frac{\Lambda}{6}(r^2-r^2_{\rm ta})\right]}}\;,
\end{equation}
and \citep{DelPopolo1999}
\begin{equation}\label{eqn:temp1}
 t_{\rm ta}=\int_0^{a_{\max}}\frac{da}{\sqrt{H_i^2\left[\frac{1+\overline{\delta}}a-
            \frac{1+\overline{\delta}}{a_{\max}}\right] + 
            \int_{a_{\max}}^a\frac{8G^2L^2}{H_{\rm i}^4r_{\rm i}^{10}(1+\overline{\delta})^2}}a^3
            +\frac{\Lambda}{6}(a^2-a_{\rm max}^2)}\;.
\end{equation}

Using Eqs.~(\ref{eqn:collss}) and (\ref{eqn:temp}) (or Eqs.~(\ref{eqn:secss}) and (\ref{eqn:temp1})) it is possible to 
obtain the linear overdensity at turnaround and collapse ($\delta_{\rm c}$), similarly to \cite{Bartlett1993}. 
Solving Eq.~(\ref{eqn:temp}) for a given mass and turn-around time gives the turn-around radius, which is related to 
the binding energy through Eq.~(\ref{eqn:collss}) with $dr/dt=0$. The binding energy of a growing mode is uniquely 
given by the overdensity $\delta_{\rm i}$ at time $t_{\rm i}$. Linear theory can be used to get the overdensity at 
turn-around and collapse, $\delta_{\rm c}$. The connection between the binding energy, $C$, and $\delta_{\rm i}$ can be 
obtained by means of the relation $v$-$\delta_{\rm i}$ for the growing mode 
\citep[e.g.,][]{Peebles1980,Bartlett1993}.

As shown in \cite{DelPopolo2006c}, the threshold becomes
\begin{align}
 \delta_{\rm c} & = \delta_{\rm co} \left[1+
                    \int_{r_{\rm i}}^{r_{\rm ta}}\frac{r_{\rm ta}L^2 \cdot {\rm d}r}{G M^3 r^3} + 
                    \Lambda \frac{r_{\rm ta}r^2}{6 G M}\right] \nonumber \\
           & \simeq \delta_{\rm co} \left[1+\frac{\beta}{\nu^{\alpha}}+\frac{\Omega_{\Lambda}\beta_2}{\nu^{\alpha_2}}
                    \right]\;,
\end{align}
where the constants were already given in section~(\ref{sect:barrier}) for Eq.~\ref{eqn:barrierl}.

For $\eta\neq 0$, the barrier can be obtained similarly to the case $\eta=0$ starting from Eq.~(\ref{eqn:coll}) 
(or \ref{eqn:sec}).

In this case, the threshold becomes
\begin{align}
 \delta_{\rm c} & = \delta_{\rm co} \left[1+\int_{r_{\rm i}}^{r_{\rm ta}} 
                    \frac{r_{\rm ta} L^2 \cdot {\rm d}r}{G M^3 r^3} + \frac{\lambda_{o}}{1-\mu(\delta)}+
                    \Lambda \frac{r_{\rm ta} r^2}{6 G M}\right] \nonumber \\
           & \simeq \delta_{\rm co} \left[1+\frac{\beta}{\nu^{\alpha}}+\frac{\Omega_{\Lambda}\beta_2}{\nu^{\alpha_2}}+ 
                    \frac{\beta_3}{\nu^{\alpha_3}}\right]\;,
\end{align}
where $\alpha_3=0.07$, $\beta_3=0.45$, $\mu(\delta) = \frac{\sqrt{2}\pi}{3c(\overline{\delta})}
\left(\frac{1}{\overline{\delta}}+1\right)^{3/2}$ is given in \cite{Colafrancesco1995} (Eq. 29); 
$\lambda_{o}=\epsilon_oT_{co}$, where
\begin{equation}
 \epsilon_o = \eta a^{3/2}=\frac{4.44[G m_a n_{ac}]^{1/2}}N\log{\left(1.12N^{2/3}\right)}\;,
\end{equation}
given $m_a$ and $n_a$, the mass and the number density of field particles (particles generating the fluctuating field), 
respectively, $N=\frac{4\pi}3R_{\rm sys}^3n_a$ their total number, $n_{\rm ac}=n_a\times a^3$ their comoving number, 
and $R_{\rm sys}$ the system radius \citep{AntonuccioDelogu1994},\citep[][Appendix D]{DelPopolo1999}, while $T_{co}$ 
is the collapse time of a pure top hat model \citep{Gunn1972}
\begin{equation}
 T_{c0}(r,\nu)=\frac{\pi}{H_{\rm i}[\overline{\delta}(r,\nu )]^{3/2}}\;.
\end{equation}

The angular momentum $L$, due to the tidal interaction with neighbours, is calculated getting the r.m.s. of the torque 
\citep[see][]{Ryden1988},\citep[][Eq.~(C5)]{DelPopolo2009}, and then integrating the torque over time 
\cite[][Eq.~(35)]{Ryden1988},\cite{DelPopolo2009} obtaining
\begin{equation}\label{eqn:ang}
 L(r,\nu) = \frac{1}{3}\left(\frac{3}{4}\right)^{2/3}\tau_ot_0\overline{\delta}_o^{-5/2} 
            \int_0^\pi \frac{\left(1-\cos{\vartheta}\right)^3}
            {\left(\vartheta -\sin{\vartheta}\right)^{4/3}}
            \frac{f_2(\vartheta)}{f_1(\vartheta)-f_2(\vartheta)\frac{\delta_o}{\overline{\delta}_o}}
            d\vartheta\;,
\end{equation}
where $\tau_o$ is the tidal torque at $t_o$, and the functions $f_1(\vartheta)$ and $f_2(\vartheta)$ are given by 
\cite{Ryden1988}, \cite[see][]{DelPopolo2009}.

\section{Expressions for the mass function}\label{sec:MF}
In the following, we write the MF using the same notations and approximations of \cite{Klypin2011}. 
The mass function (Eq.~(\ref{eqn:universal})) is given by
\begin{equation}\label{eqn:universal1}
 n(M,z)=\frac{\overline{\rho}}{M^{2}}\frac{d\log{\nu}}{d\log M}\nu f(\nu)\;.
\end{equation}
Recalling that $dn = n dM$, that the background density $\overline{\rho}=\rho_{\rm cr}\Omega_{\rm m}$ and using the 
mass function of \cite{Jenkins2001}, $f(\sigma)= \frac{M}{\overline{\rho}}\frac{dM}{d\log{\sigma^{-1}}}$, 
Eq.~(\ref{eqn:universal1}) can be written as
\begin{align}
  M\frac{dn}{dM} & = \overline{\rho}\frac{d\sigma(M)}{\sigma(M)dM} f(\sigma) = 
                     \Omega_{\rm m,0}\rho_{\rm cr,0}\frac{d\sigma(M)}{\sigma(M)dM}f(\sigma)\;,\\
                 & = 2.75\times 10^{11}(h^{-1}{\rm Mpc})^{-3}\Omega_{\rm M,0}h^2M_{\odot}
                     \frac{d\sigma}{\sigma dM}f(\sigma)\;,
\end{align}
where $\rho_{\rm cr,0}$ is the critical density today ($\overline{\rho}$ in Eq. (\ref{eqn:universal})), $M$ is the 
halo virial mass and
\begin{equation}
 \sigma^2(M) = \frac{\delta^2(a)}{2\pi^2}\int_0^\infty k^2P(k)W^2(k,M)dk\;,
\end{equation}
is the mass variance, $P(k)$ the primordial power spectrum of perturbations, $W(k,M)$ the Fourier transform of the 
real-space top-hat window function. The linear growth-rate function $\delta(a)$ is given by
\begin{equation}\label{eqn:delta}
 \delta(a) = D(a)/D(1)\;,
\end{equation}
where the expansion parameter is connected to redshift through $a=1/(1+z)$. The growth rate factor $D(a)$ can be 
approximated, for $\Omega_{\rm m}>0.1$, using the expression proposed by \cite{Lahav1991,Carroll1992}.%
\footnote{For $\Omega_{\rm M,0}=0.27$ the error is smaller than $7\times 10^{-4}$.}

Since we are going to compare our results to the Bolshoi simulation, we assume the approximation for $\sigma(M)$, the 
r.m.s. density fluctuation 
\begin{equation}
 \sigma(M) = \frac{16.9y^{0.41}}{1+1.102y^{0.20}+6.22y^{0.333}}\;,\quad 
 y \equiv \left[\frac{M}{10^{12}~h^{-1}~M_{\odot}} \right]^{-1}\;,
\end{equation}
whose accuracy is better than 2\% for $M > 10^7~h^{-1}~M_{\odot}$.

For the mass function of \cite{Sheth2002}, we have
\begin{equation}
  f(\sigma) = A\sqrt{\frac{2a_1}{\pi}}\left[1+(a_1\nu)^{-0.3}\right]\sqrt{\nu}\exp{\left({-\frac{a_1 \nu}{2}}
             \right)}\;,
\end{equation}
where $\nu \equiv \left(\frac{1.686}{\sigma(M)}\right)^2$, $A = 0.322$ and $a_1 = 0.707$.

For our barrier 
\begin{align}\label{eq:multiplic}
 \nu f(\nu) \simeq & A_2 \left(1+\frac{0.1218}{\left(a\nu\right)^{0.585}}+\frac{0.0079}{\left(a\nu\right)^{0.4}}+
                     \frac{0.1}{\left(a\nu\right)^{0.45}}\right) \sqrt{\frac{a\nu}{2\pi}} \times \nonumber\\
                   & \exp{\left\{-0.4019 a\nu^{2.12}\left[1+\frac{0.5526}{\left(a\nu\right)^{0.585}}+
                     \frac{0.02}{\left(a\nu\right)^{0.4}}+\frac{0.07}{\left(a\nu\right)^{0.45}}\right]^{2}\right\}}\;,
\end{align}
where $A_2=0.965$.

\end{document}